\documentclass[12pt]{article}
\pdfoutput=1

\usepackage{amsmath,amssymb,amscd}
\usepackage{listings}
\usepackage{caption}
\usepackage{dsfont}
\usepackage{slashed}
\usepackage{color}
\usepackage{ulem}
\usepackage[pdftex]{graphicx}
\usepackage{epstopdf}
\usepackage{subfigure}
\usepackage{epsfig}
\usepackage{cite}
\usepackage{multirow}
\usepackage{lineno}
\usepackage{comment}
\usepackage{array}

\newcommand{\beq}{\begin{equation}}
\newcommand{\eeq}{\end{equation}}

\begin{document}

\baselineskip=21pt

\begin{center}
{\Large {\bf Higgs Field-Induced Triboluminescence in Binary Black Hole Mergers}}
\vskip 0.5cm
{\bf Mariam Chitishvili}\textsuperscript{a},~
{\bf Merab Gogberashvili}\textsuperscript{a,b},~
{\bf Rostislav Konoplich}\textsuperscript{c}, ~\\
{\bf Alexander S. Sakharov}\textsuperscript{c,d}~\\
\vskip 0.5cm
{\small {\it
\textsuperscript{a}{\mbox Department of Exact and Natural Sciences, Javakhishvili Tbilisi State University}\\
\mbox{Tbilisi 0179, Georgia}\\
\vspace{0.25cm}
\textsuperscript{b} {\mbox Department of High Energy Physics, Andronikashvili Institute of Physics}\\
\mbox{Tbilisi 0177, Georgia}\\
\vspace{0.25cm}
\textsuperscript{c} {\mbox Physics Department, Manhattan College}\\
{\mbox 4513 Manhattan College Parkway, Riverdale, NY 10471, United States of America}\\
\vspace{0.25cm}
\textsuperscript{d} {\mbox Experimental Physics Department, CERN, CH-1211 Gen\`eve 23, Switzerland}
}}
\end{center}

\vskip 1.0cm


\begin{abstract}

\noindent We conjecture that the Higgs potential can be significantly modified when it is in close proximity to the horizon
of an astrophysical black hole, leading to the destabilization of the electroweak vacuum.
In this situation, the black hole should be encompassed by a shell consisting of a ``bowling substance'' of the
nucleating new phase bubbles. In a binary black hole merger, just before the
coalescence, the nucleated bubbles can be prevented from falling under their seeding horizons,
as they are simultaneously attracted by the gravitational potential of the companion. For a short time,
the unstable vacuum will be ``sandwiched'' between two
horizons of the binary black hole, and therefore the
bubbles may collide and form micro-black holes, which will be rapidly evaporated by thermal emission
of Hawking radiation of all Standard Model spices. This evaporation, being triggered by a gravitational
wave signal from the binary black hole merger, can manifest itself in observations
of gamma rays and very high energy neutrinos, which makes it a perfect physics case for
multi-messenger astronomical observations.

\vskip 5mm

\end{abstract}
\vskip 5mm

\leftline{June 2023}

\vskip 10mm

\section{Introduction}

The discovery of the Higgs boson at the LHC~\cite{higgsCMS, higgsATLAS}
was a confirmation of the Standard Model (SM) and it provided an exciting
opportunity to understand properties of the electroweak (EW) vacuum.
In particular, it was demonstrated ~\cite{higgsStrumia1, higgsStrumia2} that there is an available
lower energy vacuum state to which the EW vacuum can eventually decay. Actually, this
possibility has been known for a long time and comprehensively analyzed
in~\cite{smInstable1, smInstable2, smInstable3, smInstable4, smInstable5}. It appears that
the Higgs potential is sensitive to the experimental inputs, in particular to the physical masses
for the Higgs boson and the top quark and also to physics beyond the SM. The recent measurements
of the Higgs boson and top quark masses, $M_h=125.25\pm 0.17$~GeV,
$M_t=172.76\pm 0.30$~GeV~\cite{PDG} implies that our universe resides in the unstable SM vacuum state. Thus, if
the SM is valid up to energies greater than about $10^{12}$~GeV, the EW vacuum is
meta-stable and the transition into a lower energy state will occur in the future. The transition
happens initially locally, nucleating a small bubble of the true vacuum.
The bubble then starts expanding at high rate reaching quickly almost the speed of light and converting the meta-stable
vacuum to the true one everywhere.

Although the fact of our existence testifies that at the present day vacuum
decay rate is extremely low, this was not necessarily the case in the early universe. For example, a high Hubble rate
during inflation and high temperatures afterwards could potentially
increase the rate significantly~\cite{cosmology1, cosmology2}. The decay rate also can be very sensitive to a
presence of new physics, such as an extension of the SM  via introduction
of higher dimension operators~\cite{bsm1, bsm2}, or the Higgs potential dependence on the space-time curvature
through the direct non-minimal coupling of Higgs field to curvature
(see, for example~\cite{curvature}). The presence of a small black hole (BH) can catalyze vacuum decay and make it significantly
faster~\cite{hBH1, hBH2, hBH3, hBH4, hBH5, hBH6, hBH7}~\footnote{However, this claim has been questioned in~\cite{hBnoH1}.}.
Thus, the fact that we still
observe the universe in its EW vacuum state enables us to impose constraints on the cosmological history.
This includes factors like the reheating temperature and the scale of inflation, as well as beyond SM parameters,
such as coupling constants between the Higgs field and space-time curvature or higher dimension operators.

One could ask a question, whether there is another way, not based on the anthropic arguments, of using the meta-stability of the
EW vacuum to study extreme conditions or effects of new
physics in cosmology and astrophysics. Namely, would it be possible to think of some physical conditions at which vacuum
decay could occur and manifest itself as a potentially observable
phenomenon, while not leading to catastrophic consequences for the existence of the universe in its present
meta-stable state? In this paper we would like to propose an idea of such kind of phenomenon and elaborate on its
driving mechanism and observational signatures in multi-messenger astronomy.

The idea is creating physical conditions for putting the EW vacuum in a high decay rate regime within a finite
spatial volume for a certain time period. The desired conditions can be realized
in a close vicinity of the horizon of a BH of astrophysical origin,
as a result of gravitationally induced corrections to the Higgs potential.
In particular, the position and the height of
the potential barrier, which ``screens'' the metastable minimum of the
Higgs potential, can be modified so significantly that the vacuum becomes unstable already at present
temperatures within a certain distance above the horizon of the BH. Since the instability implies a high
nucleation rate of the Higgs new phase bubbles, a BH immersed into EW vacuum
will be encompassed by a thin shell consisting of a ``bowling substance''
represented by nucleating new phase bubbles surrounded with the SM vacuum. Although the nucleation is
permanently going on within the thickness of the shell, the bubbles
immediately fall under the horizon, so that an external observer of a single BH always stays in the EW
meta-stable state.

The situation can be different if we consider a binary black hole (BBH)
merger~\cite{bbhLIGO1, bbhLIGO2, bbhLIGO3, bbhLIGO4}, where BHs circle their common center. In the merger
the BHs spiral inward losing their orbital energies in the form of gravitational radiation so that their
horizons get very close to each other and finally coalescence to form a single BH.
Just before the coalescence, when the horizons of the components are close to each other, namely at a
distance of the order of the thickness of the shells, the nucleated bubbles can be
prevented from the falling under their seeding horizon, being pulled out by
the gravitational potential of the other component of the BBH merger. Thus, within
some volume ``sandwiched’’ in the gap between the approaching each other horizons,
a temporal stabilization of the process of the EW vacuum conversion
into a new phase can occur. Inside this volume, within the time of its existence, nucleating bubbles will expand,
collide and even percolate. Some of these collisions can result in formations of
microscopic black holes ($\mu$BHs) via mechanisms described in~\cite{pbhBubble4, pbhBubble3}.
Namely, when three bubbles collide the surface energy in parts of their walls can be focused to
the extent that its density tends to infinity, which converts the triple collision point into a BH~\cite{pbhBubble4}.
Moreover, the collapse of non-trivial vacuum structure left over
after collisions of only two bubbles can also lead to the formation of a BH, as it is argued
in~\cite{pbhBubble3,pbhBubble6}~\footnote{See also a recent study of the mechanism~\cite{pbhBubble3} presented in~\cite{pbhBubble5}.}.
The masses of such a ``split off'' from the merger $\mu$BHs depend on
the sizes of the colliding bubble and their walls tension. The sizes of the
bubbles at the collisions are mostly determined by the phase transition developed in the volume of
the stabilized SM vacuum decay process. The wall tension should depend on the details of modification of the position of
the barrier separating the meta-stable EW vacuum from the true
vacuum state and should be driven by the energy scale not essentially exceeding the EW one.

The formed $\mu$BHs will start to evaporate emitting thermal Hawking
radiation~\cite{Hawking} in all SM spices. Finally, the coalescence of the merger results in the formation of a single
horizon of the final BH. Thus, the conditions of vacuum instability
created in the volume ``sandwiched'' in the gap between the horizons
of the merger are broken so that the stabilized conditions for bubble nucleation are destroyed.
All remnants of new vacuum phase and still not evaporated small $\mu$BHs should fall into the final BH. This ensures
that an external observer will remain in the EW vacuum state.

The duration of the emission of Hawking radiation by the $\mu$BHs cannot exceed
the typical time
scale of the last portion of gravitational wave signals
from BBH mergers discovered by LIGO and
Virgo~\cite{bbhLIGO1, bbhLIGO2, bbhLIGO3, bbhLIGO4}, which is much
less than a second. One might expect that the energy release of
Hawking radiation can be at the level of isotropic energy equivalents
measured for short gamma ray bursts (SGRBs)~\cite{SGRB}. Therefore,
the electromagnetic part of the burst could be observed by space based gamma ray bursts monitors~\cite{SWIFT, INTEGRAL,
Fermi-GBM} and the telescope~\cite{Fermi-LAT}. Moreover, the Hawking emission of other SM model particles  and,
maybe, beyond SM spices can produce a neutrino signal in IceCube~\cite{IceCube1}.
The very high energy spectral part of the electromagnetic component of the
Hawking radiation also might produce a signature in very high energy atmospheric Cherenkov
facilities~\cite{MAGIC, HESS, VERITAS, HAWC, LHAASO}.

The mechanism described above is akin to so called triboluminescence~\cite{tribo}
which refers to the phenomenon that materials could emit light when they are mechanically
stimulated, such as rubbing, grinding, impact, stretching, and compression. Here, with a certain degree of analogy,
the vacuum ``bowling substance'', being the ``material'' of
the shell encompassing the horizon of an astrophysical BH,
is getting stretched between two approaching each other horizons in a
BBH merger which finally leads to an emission of detectable
Hawking radiation.

The paper is organized as follows. In Section \ref{sec:sandw} we describe heuristically general phenomenological
features of the proposed mechanism of the formation and evaporation of
$\mu$BHs within the gap between horizons of a BBH merger components. In Section \ref{sec:emSignal}
we study basic features of spectra, temporal characteristics and energy budget of the
electromagnetic and neutrino messengers of the phenomenon. In Section \ref{sec:SMvacuumSt}
we review the Higgs field effective potential in the Standard Model; in Sections
\ref{sec:GravCorr} and \ref{sec:QvacuumH} the idea of the mechanism of gravitationally
induced corrections to the Higgs potential in a vicinity of a BH horizon is introduced. In Section
\ref{sec:ToyM} estimates of basic quantities driving the proposed mechanism are performed in the
framework of a toy model. The energy budget of the Higgs induced triboluminescence phenomenon,
in the framework of the toy model, is discussed in Section \ref{sec:disc}.
Finally, conclusions are presented in Section \ref{concl}.


\section{Formation of $\mu$BHs in unstable vacuum ``sandwich''}
\label{sec:sandw}

To specify further the anticipated
multi-messenger manifestations of the phenomenon outlined in the
introduction, it is helpful to provide a heuristic sketch of the
underlying mechanism of gravitational corrections to the vacuum
decay process. A more detailed description of the Higgs phase
transition at nucleation sites located
in the vicinity of the horizon will
be deferred to subsequent sections.

The rate of the first order phase transition with Higgs field,
when its nucleation site is located at the center of a BH, has been
investigated in~\cite{hBH1, hBH2, hBH3, hBH4, hBH5, hBH6}.
In these studies, it was concluded that Higgs true vacuum formation rate can be increased so much that even a
single BH of smallest mass ($\ll 1$~g) existed in the past could already
destroy the current SM meta-stable vacuum state. Hence, it might happen
that any primordial BH with a lifetime smaller than the age of the universe
could serve as a source of such a fatal catalysis corresponding to
Higgs vacuum phase transition.

Here we conjecture the Higgs phase transition rate also could be substantially increased in the vicinity of the horizon of a BH,
while the nucleation site is located outside the BH. Some recent
efforts to investigate such a possibility have been made in~\cite{hBHout1, hBHout2}.
Heuristically thinking, one presumes that the closer is a nucleation site to the horizon of a BH the higher
is the phase transition rate and hence the higher is the probability
of formation of a bubble of the true Higgs vacuum. Moreover, one might expect that a BH of any mass, even such
as that one of an astrophysical origin, could create a nucleation site in
the vicinity of its horizon. In this case, since the nucleation site with a bubble of the true vacuum falls quickly
under the horizon, it is obvious that the fatal catalysis of the Higgs vacuum phase transition cannot take place.

The bubble formation probability is driven by $\exp [-S_4]$, where $S_4$ stands for the four-dimensional
Euclidean action computed along the tunneling trajectory for the spherical bubble
solution~\cite{Coleman:1977, bubbleOkun1,Coleman:1980aw}. In general,
gravitational correction effects in the vicinity of a BH depend on the distance to the horizon,
so that $S_4$ can be represented as an effective action $S_4(d_H)$ which
implies that the probability of tunneling per unit time per unit volume from a
vacuum in the meta-stable state to the true vacuum is given by
\begin{equation}
\label{PhTrR1}
{\Gamma (d_{\rm H})} = {\cal M}^4\exp{[-S_{4}(d_{\rm H})]}~,
\end{equation}
where the pre-factor ${\cal M}$ is of mass dimension and $d_{\rm H}$ is
the distance to the horizon measured in units of the Schwarzschild radius $R_{\rm S}$ of the BH. For illustration, in the
regime of gravitational corrections, one may model the effective action with a simple power law
\begin{equation}
\label{effAct1}
S_{4}(d_{\rm H} ) = A_{\rm S}d_{\rm H}^{a}~,
\end{equation}
where $A_{\rm S}$ indicates the value of the action at the distance
$d_{\rm H}= 1$, that is to say at one Schwarzschild radius from the
horizon and $a$ is the power to be specified from the details of
the gravitational corrections for the Higgs potential~\footnote{Choosing (\ref{effAct1}),
we were governed by the argument that power laws can naturally satisfy the demand
that a distance dependence of the nucleation rate is invariant under arbitrary
rescaling, up to normalization measured here by the Schwarzschild radius.}.
Therefore, the closer is the location of a bubble nucleation site to the horizon
the higher is the probability of tunneling from the
meta-stable vacuum state at $v=246$~GeV to the true Higgs vacuum.
If one goes away from the horizon, the exponential suppression in (\ref{PhTrR1}) slows down the bubble
nucleation rate so that they would have more time to grow up before starting
to collide if they were prevented from falling down into the BH. This effect could take place in a BBH
merger consisted of BHs of about LIGO-Virgo scale~\cite{bbhLIGO1, bbhLIGO2, bbhLIGO3, bbhLIGO4}, just before touching
each other by their horizons.

Indeed, between the horizons of the BBH components, at their close enough mutual approach,
one might expect an instant formation of a volume with effective zero gravity so that
for a short time period the unstable vacuum is kind of ``sandwiched'' between the horizons
of the components. In these conditions bubbles could have enough time to grow up and
finally percolate, terminating their existence in wall collisions.
In triple bubble collisions the surface energy of the parts of the colliding walls can be focused in a way that they find
themselves under their Schwarzschild radius leading to a
formation of $\mu$BHs, as argued in~\cite{pbhBubble4}. The formation of $\mu$BHs is also possible
through evolution of non-trivial vacuum configuration produced in a
collision of walls of only two bubbles~\cite{pbhBubble3,pbhBubble6}. Either case is feasible in conditions of
percolation within quite a limited volume. Being formed, the $\mu$BHs
will be rapidly evaporated by a thermal emission of Hawking radiation~\cite{Hawking} with black-body spectrum at an effective
temperature which increases as the mass of a $\mu$BH decreases.

In general, the masses of such ``split off'' $\mu$BHs are defined by sizes of the bubbles at the instance of
their collision and their vacuum wall tensions. The wall tension is mostly defined
by parameters of the modified Higgs potential, namely by a location of the maximum of the barrier
separating the meta-stable EW vacuum from the true vacuum state and the Higgs self-interaction
constant $\lambda$, at the barrier. Both parameters for unmodified potential are quite well defined by
the measurements of the Higgs and top masses as well as the SM physics
renormalization group corrections~\cite{cosmology2}.

The surface tension is estimated as
\begin{equation}
\label{tl2}
\sigma\simeq\sqrt{\lambda}h_{\rm b}^3~,
\end{equation}
where $h_{\rm b}$ the Higgs field value defined by a bounce solution at the modified maximum, which, for sanity reasons,
cannot substantially exceed the position of the electroweek vacuum, so
that
\beq
\label{hb1TeV}
h_{\rm b}\lesssim 1\ {\rm TeV}~.
\eeq
It is reasonable to assume that the mass of a $\mu$BH formed in a collision contains fraction $\kappa_f$
of the mass of the colliding bubbles. Thus, one can
say that a $\mu$BH of mass $M_{\mu{\rm BH}}$ is created out of a bubble of radius
\begin{equation}
\label{tl3}
R_{\rm bub}\simeq\frac{M_{\mu{\rm BH}}^{1/2}}{(4\pi)^{1/2}\lambda^{1/4}h_{\rm b}^{3/2}\kappa_f^{1/2}}
\approx\frac{l_{\rm Pl}}{2\sqrt{\pi}\lambda^{1/4}\kappa_f^{1/2}}
\left(\frac{M_{\mu{\rm BH}}}{M_{\rm Pl}}\right)^{1/2} \left(\frac{M_{\rm Pl}}{h_{\rm b}}\right)^{3/2}~,
\end{equation}
where $l_{\rm Pl}$ and $M_{\rm Pl}$ are the Planck length and Planck mass, respectively.
The value of $\kappa_f$ depends on the details of a conversion mechanism and, in general, it
is expected to amount a non-negligible (up to a per cent) fraction of the
bubble mass~\cite{pbhBubble3,pbhBubble6}.

For the sake of simplicity of the rough estimates, let us assume that the volume of the unstable vacuum ``sandwich'' (UVS)
defined by $d_{\rm H}$ is populated with bubbles
of approximately equal sizes, which implies that all ${\mu}$BHs in this volume are formed with equal masses.
In the course of the bubble nucleation and growth, the  ${\mu}$BHs
will be created in bubbles double and triple collisions, as argued above.
Therefore, one has to estimate the abundance of such collisions occurring within the volume $V_{\rm UVS}(d_{\rm H})$.

\begin{figure}
\centering
\includegraphics[angle=0,scale=0.4]{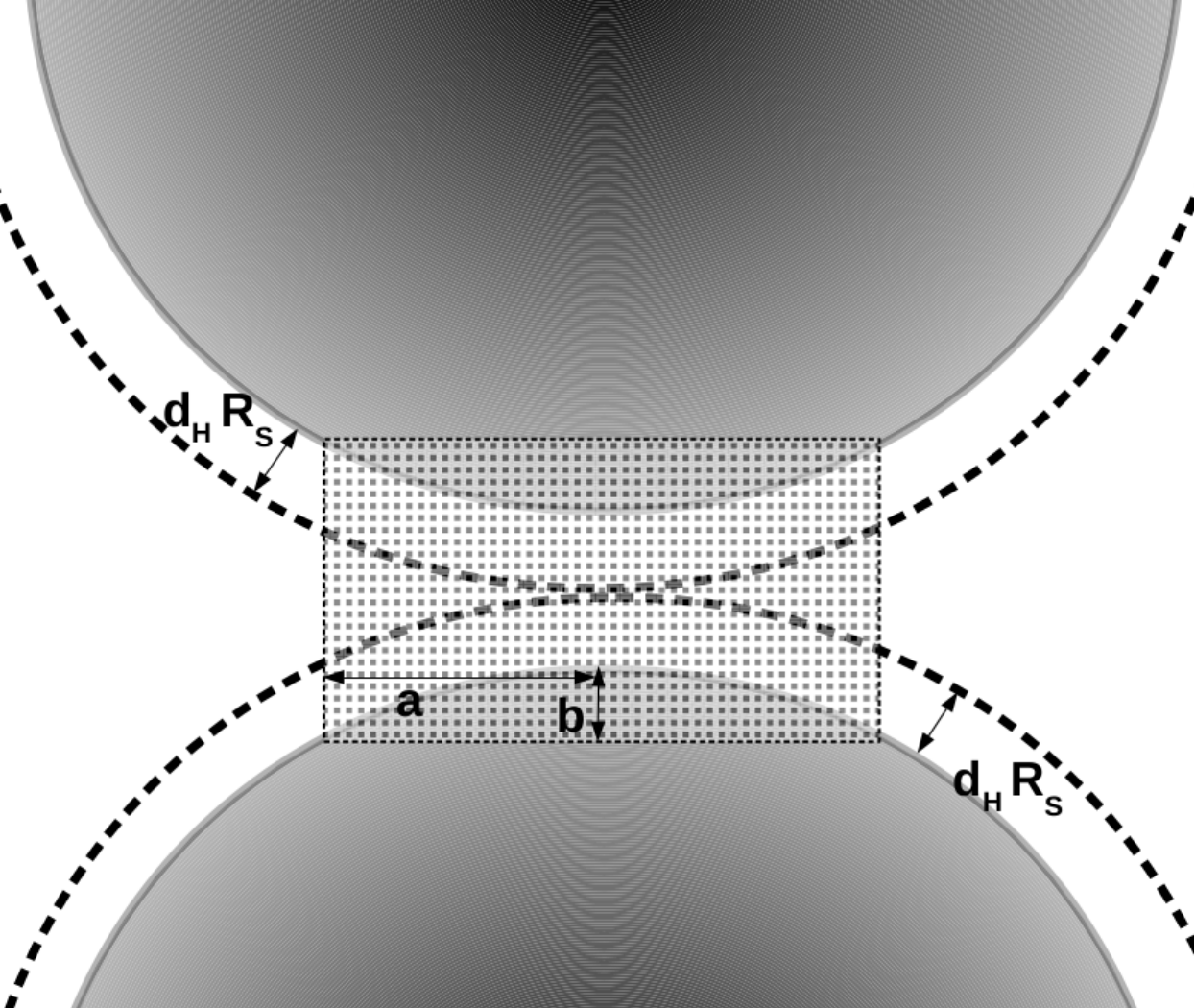}
\vspace{0.5cm}
\caption{\it A schematic representation of the UVS formed just before the horizons of the BBH components,
shown by the gray gradient parts of spheres, come into contact. The dashed enveloping line represents the
layer spreading outside the horizons of the BHs, which contains the unstable vacuum.}
\label{fig:UVS}
\end{figure}

The volume $V_{\rm UVS}(d_{\rm H})$ is assumed to be formed within a narrow gap between the
spherical caps of the horizons of the components of the BBH, just before their mutual touching,
as shown in Fig.~\ref{fig:UVS}. The thickness of the USV can be taken as $2d_{\rm H}R_{\rm S}$,
provided that the power in (\ref{effAct1}) is not very high, so that the transition rate does not change
dramatically across the UVS. Thus one can represent $V_{\rm UVS}(d_{\rm H})$ as a body bounded between the convex surfaces of two spherical caps separated, at their closest
points, by distance $2d_{\rm H}R_{\rm S}$, as shown in Fig.~\ref{fig:UVS}. The volume $V_{\rm UVS}(d_{\rm H})$  can be calculated as
\begin{equation}
\label{vol1}
V_{\rm UVS}(d_{\rm H})\approx V_{\rm cyl}(d_{\rm H})-2V_{\rm cap}(d_{\rm H})~,
\end{equation}
where $V_{\rm cyl}(d_{\rm H})$ is the cylindrical volume formed by the base circles of
radius $a$ separated by distance $3d_{\rm H}R_{\rm S}$. This insures that the width of the UVS,
at its edges framed by the caps, does not increase more than by a factor 1.5, keeping the Euclidean action (\ref{effAct1}) increasing only within factor $(1.5)^{a}$,
which is moderate, provided that, once again, the power $a$ is not very large.
Taking into account that according to the construction of the cylinder above (see Fig.~\ref{fig:UVS}),
$b=\frac{1}{2}d_{\rm H}R_{\rm S}$, the radius of the base circle is given by
\beq
\label{bcirc1}
a=\sqrt{b(2R_{\rm S}-b)}\approx\sqrt{d_{\rm H}}R_{\rm S}~,
\eeq
which allows us to express the UVS volume as~\footnote{Standard textbook formulas were used to obtain this result.}
\begin{equation}
\label{vol2}
V_{\rm UVS}(d_{\rm H})=\frac{1}{6}\pi b(3a^2+b^2)\approx \frac{5}{2}\pi d_{\rm H}^2R_{\rm S}^3~.
\end{equation}
Evaluating (\ref{vol1}), we considered BBH of equal mass components, so that both caps have the same volume.
Also, obtaining (\ref{vol2}) we ignored everywhere $d_{\rm H}$ with respect to values $\gtrsim 1$.
Taking into account these two approximations, it would be more correctly to say that (\ref{vol1})
expresses a sort of upper estimate of the UVS volume within which
the action (\ref{effAct1}) does not variate more than by factor $(1.5)^{a}$.
Thus, bellow, for simple estimates, we will use a constant value of the action (\ref{effAct1}),
defined by $d_{\rm H}$ inside the UVS.

Depending of the degree of the closeness of the horizons of the BBH components in the course of their mutual approach, two different scenarios of
bubble collisions leading to formation of ${\mu}$BHs of distinct masses are possible.

The {\it first scenario} refers to the regime when the bubble nucleation rate is significantly enhanced in comparison with the nucleation rate in the horizons free space,
while it is still low in the sense of providing the percolation of the bubbles of radius $R_{\rm bub}$ within $V_{\rm UVS}(d_{\rm H})$.
Therefore, in this scenario, the only bubbles which nucleated in closed pairs and closed triplets would collide and form ${\mu}$BHs.
In order to estimate the number of these collisions, one can use some findings from the site percolation theory~\cite{perc1}, where the random distribution of occupied and empty
sites on a lattice is considered. The occupied sites tend to aggregate into clusters of size distribution $n_s(p)$~\footnote{This quantity serves a discrete analog
of distribution function in statistical physics.},
which is defined as a number of clusters containing $s$ occupied sites per one lattice site when the
fraction of occupied sites on the lattice is equal to $p$. The cluster numbers are calculated
on the basis of lattice animals counting, and for large clusters and low occupancy ($p\rightarrow 0$) it is given by the corresponding equation in ~\cite{perc1}:
\beq
\label{pcl1}
n_s(p)\propto s^{-\theta}p^sC^s~,
\eeq
where $C$ is a constant. For a three dimensional lattice, $\theta = 3/2$, as inferred in~\cite{perc2}.
Similar behavior as (\ref{pcl1}) has been confirmed for small clusters as well in simulations
performed in~\cite{perc3}. In the context of the formation of ${\mu}$BHs in
bubbles collisions, we are interested in clusters of size $s=2$ and $s=3$ aggregated
on a simple 3D lattice of volume $V_{\rm UVS}(d_{\rm H})$ with sites measured by the
volume of individual bubble $V_{\rm bub}$ defined, in its turn, by the radius (\ref{tl3}).
Therefore the number of ${\mu}$BHs created inside $V_{\rm UVS}(d_{\rm H})$
out of collisions of bubbles of volume ${V_{\rm bub}}$ can be expressed as
\beq
\label{Npc1}
N_{\mu{\rm BH}}(p,s)\approx n_s(p){\cal P}_{\mu{\rm BH}}(s)\frac{V_{\rm UVS}(d_{\rm H})}{V_{\rm bub}}~,
\eeq
where $s=2,3$ and ${\cal P}_{\mu{\rm BH}}(s)$ is the probability of formation of the ${\mu}$BH
by two~\cite{pbhBubble3} or three~\cite{pbhBubble4} bubbles.
The fraction of occupied sites on this lattice can be expressed as
\beq
\label{Npc2}
p(d_H)\simeq \langle \Gamma (d_{\rm H}) \rangle \Delta t_{\rm col}V_{\rm UVS}(d_{\rm H})~,
\eeq
where $\langle \Gamma (d_{\rm H})\rangle$ is the nucleation rate averaged over $V_{\rm UVS}(d_{\rm H})$ and
$\Delta t_{\rm col}$ is the time elapsed from the bubbles nucleation to their collisions. For the small
clusters number we use the result obtained in~\cite{perc3} for a cubic lattice, which reads
\beq
\label{Npc3}
n_s(p)\approx s^{-1.5}p^s10^{1.1s},
\eeq
where the numerical values for $\theta$ and $C$ have been inferred reading out
the relevant parameters from Fig.5 of~\cite{perc3}. Thus, collecting together (\ref{Npc1}), (\ref{Npc2}) and (\ref{Npc3}),
the number of ${\mu}$BHs created in the UVS can be represented as
\begin{equation} \label{tl4}
N_{\mu{\rm BH}}(d_H,s)\approx s^{-1.5}10^{1.1s}\langle \Gamma (d_{\rm H}) \rangle^s \Delta t_{\rm col}^s{\cal P}_{\mu{\rm BH}}(s)
\frac{\left[V_{\rm UVS}(d_{\rm H})\right]^{(s+1)}}{V_{\rm bub}}~.
\end{equation}
The collision time in (\ref{tl4}) can be evaluated as the time spent by bubbles
while they were growing from the critical radius $\rho_c$ up to ${R_{\rm bub}}$ given in (\ref{tl3}),
which implies that
\begin{equation}
\label{t_pc1}
\Delta t_{\rm col}\approx\frac{R_{\rm bub}}{c}~,
\end{equation}
where $c$ is the velocity of light.

It is practical to express (\ref{tl4}) separately for $s=2$ and $s=3$ mechanisms as follows
\begin{equation}
\begin{split}
\label{nmbh1}
N_{\mu{\rm BHs}}(d_{\rm H},2)\approx 1.6\times 10^{31}d_H^6\lambda^{1/4} \langle \Gamma (d_H) \rangle^2 {\cal P}_{\mu{\rm BH}}(2)(R_{\rm S\odot})^9\kappa_f^{1/2}\times \\
\qquad\qquad\times\left(\frac{M_{\rm BH}}{M_{\odot}}\right)^{9} \left(\frac{M_{\rm Pl}}{M_{\mu{\rm BH}}}\right)^{1/2}
\left(\frac{h_{\rm b}}{M_{\rm Pl}}\right)^{3/2}\left(\frac{\rm 1\ s^2}{\rm 1\ km^3}\right)~,
\end{split}
\eeq
\beq
\label{nmbh2}
N_{\mu{\rm BHs}}(d_{\rm H},3)\approx 1.3\times 10^{-11}d_H^8\langle \Gamma (d_H) \rangle^3 {\cal P}_{\mu{\rm BH}}(3)(R_{\rm S\odot})^{12} \left(\frac{M_{\rm BH}}{M_{\odot}}\right)^{12}
\left(\frac{\rm 1\ s^3}{\rm 1\ km^3}\right)~,
\eeq
where $R_{\rm S\odot}=2.95$~km stands for the Schwarzschild radius of a solar mass object.
Since dynamical effects, which might be caused by gravitational forces applied at the sites of bubbles collisions are not taken into account,
the ${\mu}$BHs counted by (\ref{nmbh1}) and (\ref{nmbh2}) are assumed to be scattered uniformly across the $V_{\rm UVS}(d_{\rm H})$.

The {\it second scenario} refers to the regime when one can allocate a volume, fitting within $V_{\rm UVS}(d_{\rm H})$, where the nucleation rate
reached the high enough level to make the bubbles of radius $R_{\rm bub}$ percolating. This regime implies that, in fact, all bubbles
participate in both double and triple collision mechanisms of ${\mu}$BHs formation. The necessary condition for realization
of such a scenario is quite simple, namely the probability of nucleation of a bubble of critical radius $\rho_c$ (critical volume $V_c$) within the collision time
 $\Delta t_{\rm col}$ should be close to unity. The latter implies the validity of the following relation
\beq
\label{condSc2}
\langle \Gamma (d_H) \rangle\Delta t_{\rm col}V_{\rm c}\simeq 1~,
\eeq
where the critical radius is defined by the mechanism of the correction of the Higgs potential and
will be estimated in Section \ref{sec:ToyM}. Unlike in the first scenario, here, the number of
created ${\mu}$BHs cannot be calculated directly but it can be normalized by the total energy output
to be estimated in the next section.

It is obvious, that in reality, while the horizons of components in the BBH merger are
approaching each other, one has to expect a mixture of contributions from
both scenarios. However, this complication does not affect the estimate presented
below, so that one can postpone its discussion for later publications.


\section{The electromagnetic and neutrino signals}
\label{sec:emSignal}

In this section we consider possible observational signatures of the Hawking evaporation of the ${\mu}$BHs,
which is the essence of the multi-messenger nature of the phenomenon described above.

The luminosity of gravitation signal from a BBH merger is described by the quadrupole formula
\begin{equation} \label{lumQ1}
L_{\rm GW}=\frac{1}{5}G\left<\dddot I_{jk}\dddot I_{jk}\right>,
\end{equation}
where $\dddot I_{jk}$ is the third time derivative of the quadrupole moment of a relevant mass distribution.
For a binary system with similar mass components $M_1\simeq M_2\simeq M$ separated by $R_{\rm BBH}$, the quadrupole moment can be estimated as
\begin{equation} \label{quadr1}
I\sim MR_{\rm BBH}^2~.
\end{equation}
Its third derivative is proportional to the third power of the orbital angular frequency
\begin{equation} \label{Omega1}
\Omega =\sqrt{\frac{GM}{R_{\rm BBH}^2}}~,
\end{equation}
so that
\begin{equation} \label{quadr2}
\dddot I\sim\Omega^3MR_{\rm BBH}^2~.
\end{equation}
Therefore the luminosity (\ref{lumQ1}) can be expressed as
\begin{equation} \label{lumQ2}
L_{\rm GW}\sim G\Omega^6M^2R_{\rm BBH}^4=\frac{G^4M^5}{R_{\rm BBH}^5}~.
\end{equation}
From the other hand, the gravitational energy of such BBH to be released by a merger can be expressed as
\begin{equation} \label{energy1}
E_{\rm GW}=\frac{GM^2}{R_{\rm BBH}}~.
\end{equation}
Therefore, the time scale of the merger is defined as
\begin{equation}
\label{mergeT1}
t_{\rm mrg}\simeq\frac{E_{\rm GW}}{L_{\rm GW}}\simeq\frac{R_{\rm BBH}^4}{G^3M^3}~.
\end{equation}
At the final stage of the merger the distance between the components contructs down to their Schwarzschild radius, $R_{\rm S}=2GM$,
so that the time scale (\ref{mergeT1}) of this stage is measured in $t_{\rm mrg}\gtrsim 1$~ms. Note that a BBH composed
of stellar mass BHs localized at separation distance about 10-100 times of their Schwarzschild radii would radiate out
its gravitational energy within an hour time scale. While a BBH with $\simeq 10^3$ times larger orbit will need
a Hubble time scale to exhaust its energy into emission of gravitational waves.

In terms of the gravitational wave signal the final merger stage corresponds to the gravitational wave frequency $f_{\rm max}$
at which the waveform has a maximal amplitude, as it is shown in Fig.~1 and Fig.~2 of~\cite{bbhLIGO1} for GW150914,
as well as in Fig.~10 of~\cite{bbhLIGO2} for frequency maps and reconstructed signal waveforms for other BBH events.
Using these figures we read out the conservative low values of the maximal frequencies $f_{\rm max}\simeq 100$~Hz,
which indicates that the horizons of the BHs in a BBH spent about $t_{\rm mrg}\approx 10$~ms in the closest vicinity
from each other. We treat this time scale as a rough estimate of the upper limit of the lifetime of the UVS.

We assume that during their stabilized existence period the vacuum bubbles will be converted into $\mu$BHs by
the mean of the mechanisms described in Section~\ref{sec:sandw}, which
evaporated emitting SM particles with black body spectrum~\cite{pbhEvap},
\begin{equation} \label{bhSP1}
\frac{d^2N}{dEdt}=\frac{\Gamma_s}{2\pi\left(\exp\left(\frac{E}{T_{\mu{\rm BH}}}\right)-(-1)^{2s}\right)}~,
\end{equation}
characterized by the Hawking temperature
\begin{equation} \label{bhT1}
T_{\mu{\rm BH}}=\frac{1}{8\pi}M_{\rm Pl}\left(\frac{M_{\rm Pl}}{M_{\mu{\rm BH}}}\right) \approx 10^{5}\left(\frac{10^{8}{\rm g}}{M_{\mu{\rm BH}}}\right)\ {\rm GeV}
\end{equation}
and absorption probability $\Gamma_s$. Provided that Hawking evaporation time is given by
\begin{equation} \label{bht1}
t_{\rm ev}=\frac{5120\pi}{M_{\rm Pl}}\left(\frac{M_{\mu{\rm BH}}}{M_{\rm Pl}}\right)^3 \approx 0.084\left(\frac{M_{\mu{\rm BH}}}{10^{8}{\rm g}}\right)^3\ {\rm s}~,
\end{equation}
one can conclude that $\mu$BHs with mass $M_{\mu{\rm BH}}\approx 5\times 10^7$~g will be completely evaporated
out within the time scale of $t_{\rm mrg}\approx 10$~ms emitting SM species at energies above 200~TeV, as it follows from (\ref{bhT1}).

Photons of energy $E_{\gamma}$ propagating through a background of soft photons of energy $\epsilon$ wiil produce $e^+e^-$ pairs,
if their energies exceed the threshold~\cite{nikishov, gEBL1}
\begin{equation}
\label{thr1}
E_{\gamma}\ge\frac{m_e^2}{\epsilon}\simeq 260\left(\frac{\epsilon}{1\ {\rm eV}}\right)^{-1}\ {\rm GeV}~.
\end{equation}
Soft photons with energies from 0.1 to 10~eV being produced by star formation in galaxies are abundant in the universe and known as the extragalactic background light
 (EBL)~\cite{gEBL2, gEBL3}. The profile of the spectral energy density of EBL contains two bumps located at the near infra-red energy $\epsilon\simeq 1$~eV,
 formed due to the direct starlight emission, and the far-infrared energies $\epsilon\simeq 10^{-2}$~eV, produced by scattering of starlight on dust.
 The energy density of the EBL is estimated to be about $\rho_{\rm EBL}\simeq 10^{-2.5}\ {\rm eV/cm^3}$~\cite{gEBL4}, which is a factor $\simeq 10^{-2}$
 below than the energy density of CMB. Photons with energies above 100~GeV should interact with EBL, which leads to an energy-dependent suppression of
 their flux from extra-galactic sources. The mean free path of such  $\gamma$-rays is given by
\begin{equation} \label{thr2}
\lambda_{\gamma\gamma}(E_{\gamma})\simeq\frac{1}{\sigma_{\gamma\gamma}n_{\rm EBL}}
\simeq 410\left[\frac{\pi r_0^2}{\sigma_{\gamma\gamma}}\right]\left[\frac{\epsilon}{1\ {\rm eV}}\right]
\left[\frac{10^{-2.5}\ {\rm eV/cm^3}}{\rho_{\rm EBL}}\right]\ {\rm Mpc}~,
\end{equation}
where $\sigma_{\gamma\gamma}$ is the $e^+e^-$ pair production cross section~\cite{berest, QED}
\begin{equation}
\label{siggg}
\sigma_{\gamma\gamma} = \pi r_0^2 x^{-1} \left[(2+2x^{-1}-x^{-2}) \ln(\sqrt x (1+\sqrt{1-x^{-1}})) - (1+x^{-1})\sqrt{1-x^{-1}}\right]~.
\end{equation}
In (\ref{siggg}) $r_0 = 2.82 \times 10^{-13} $cm is the classical electron radius, $x=s/4m_e^2$, $m_e$ is the electron mass,
and $s=2 \epsilon E_{\gamma \gamma} (1-cos\theta)$ is the squared center
of mass energy.
The cross section (30) has its maximum~\cite{gEBL5} $\sigma_{\gamma\gamma} \simeq 2r_0^2$ at $\sqrt{s} \simeq 1.44$~MeV leading
to the most effective interactions with the EBL photons of energy $ \epsilon \simeq 0.5 \times 10^{-2} ({\rm 100 TeV}/{E_{\gamma \gamma}}) \ {\rm eV}~$.

Photons created in the vicinity of a BH should experience the gravitational
red shift.  The photon of energy $E_{\gamma\rm em}$ emitted
at the radius $r_{\rm em}=R_{\rm S}(1+d_{\rm H})$ is observed at energy
\beq
\label{zrgav1}
E_{\gamma}=\frac{E_{\gamma\rm em}}{z_{\rm g}+1},
\eeq
by an external observer where red shift $z_{\rm g}$ is given by (see for example the text book~\cite{gravMTW})
\beq
\label{zgrav2}
z_{\rm g}=\frac{1}{\sqrt{|g_{00}(r_{\rm em})|}}-1=\frac{1}{\sqrt{1-R_{\rm S}/r_{\rm em}}}-1.
\eeq
Thus, the energy of an exiting photon will be scaled down by factor
\beq
\label{zgrav3}
z_{\rm g}+1=\sqrt{(1+d_{\rm H})/d_{\rm H}}\ .
\eeq
It implies that for a typical distance $d_{\rm H}\simeq 0.3$ the energy of photons detected by an external observer
$E_{\gamma}\simeq 100$~TeV would be about a half of the energy of the emitted photons because of photon propagation in
strong gravitational fields of the merging BHs.

According to (\ref{thr2}) and (\ref{siggg}) the mean free path $\lambda_{\gamma\gamma}$ of most of the
photons with energy $E_{\gamma}\simeq 100$~TeV can extend up to about 10 Mpc, if photons interact with EBL
of $\epsilon\simeq 10^{-2}$~eV, which is much less than the typical distance $D$ to any extra galactic source,
so that the source fluxes of such $\gamma$-rays are suppressed by factor $\exp (-D/\lambda_{\gamma\gamma})$.
Since the number density
of the target EBL photons increases with the cosmological red shift as $(1+z)^3$, (\ref{thr2}) gives
the overestimated upper limit of the mean free pass
$\lambda_{\gamma\gamma}({\rm 100\ TeV})\lesssim 10$~Mpc.
More rigorous calculations (see for example~\cite{cascade1} for details) show that
$\lambda_{\gamma\gamma}({\rm 100\ TeV})\lesssim 0.5 \ - 2$~Mpc for sources at red shifts between 0.5 and 0.1,
respectively. We notice that about half of BBH mergers detected in the first and second observing runs (O1 and O2) of the
Advanced gravitational-wave detector network~\cite{bbhLIGO2} and three quarter of BBH mergers seen in the first half of
third observing run (O3)~\cite{bbhLIGO3} performed by advanced LIGO and Virgo detectors are over $z\simeq 0.2$.

However, being absorbed on EBL, the very high energy $\gamma$-rays inject $e^+e^-$ pairs in the inter galactic media (IGM). These highly relativistic pairs are aligned with the
beam line of $\gamma$-rays emitted from the source, at a distance about $\lambda_{\gamma\gamma}$. Therefore, the full power of the source contained in the absorbed VHE $\gamma$-rays is transferred into the energy of the pairs. In their turn, electrons and positrons very effectively lose their energies via inverse Compton scattering on CMB photons. The distance of inverse Compton scattering energy attenuation of relativistic pairs of energy $E_e\simeq E_{\gamma}/2$ is given by~\cite{IC1}
\begin{equation} \label{DIC1}
D_{\rm IC}=\frac{3m_e^2}{4\sigma_T\rho_{\rm CMB}E_e}\simeq 0.37\left(\frac{E_e}{1\ {\rm TeV}}\right)^{-1}\ {\rm Mpc}~,
\end{equation}
where $\rho_{\rm CMB}\simeq 0.26\ {\rm eV/cm^3}$ is the energy density of the CMB and $\sigma_T$ is the Thomson cross section. The mean energy of the photons produced in the inverse Compton effect is calculated as~\cite{IC1}
\begin{equation} \label{EIC1}
E_{\rm IC}=\frac{4\epsilon_{\rm CMB}E_e^2}{3m_e^2}\simeq 3.6\left(\frac{E_e}{1\ {\rm TeV}}\right)^{2}\ {\rm GeV}~,
\end{equation}
where $\epsilon_{\rm CMB}\simeq 3T_{\rm CMB}$ stands for the mean energy of the CMB photons.

Therefore, in case of maximally efficient triboluminescence, which implies that $\mu$BHs are small enough to radiate out
their whole energy within the duration time of the final stage of the merger, one would have the following picture of
the propagation of the electromagnetic component. The primary short duration $t_{\rm mrg}\simeq 10\ {\rm ms}$
signal of thermal VHE $\gamma$-rays of energies $E_{\gamma}\simeq 100$~TeV emitted at the BBH merger will be
converted via $e^+e^-$ pair production on EBL and their subsequent inverse Compton scattering of CMB into softer,
however still VHE $\gamma$-rays of energies  $E_{\gamma{\rm IC1}}\simeq 10$~TeV, as it follows from (\ref{EIC1}).
This conversion will occur within the main free path of 100~TeV $\gamma$-rays,
$\lambda_{\gamma\gamma}({\rm 100\ TeV})\lesssim 0.5 \ - 2$~Mpc
(for sources at red shifts between 0.5 and 0.1), since the Compton scattering distance of $\simeq 5$0~TeV pairs is negligibly short (\ref{DIC1}).
In its turn, the mean free path of $\simeq 10$~TeV $\gamma$-rays,
which are mostly getting absorbed by the near infra-red part of the EBL spectrum,
amounts $\lambda_{\gamma\gamma}({\rm 10\ TeV})\lesssim 50 \ - 100$~Mpc~\cite{cascade1} (for sources at red shifts between 0.5 and 0.1),
which implies that, for a distant source, a good fraction of them  will be again converted into $e^+e^-$,
whose subsequent inverse Compton scattering on CMB will produce $\gamma$-rays of energy $E_{\gamma{\rm IC2}}\lesssim 300$~GeV.
Also, still some of photons will travel a long distance without significant energy loss. Thus, the power of
Hawking VHE photons tend to be converted into the secondary sub TeV $\gamma$-rays. For a source,
at distance larger than $\lambda_{\gamma\gamma}({\rm 10\ TeV})$ the cascade will be also populated by much softer
$\gamma$-rays of energy $E_{\gamma{\rm IC2}}\lesssim 3$~GeV. As it seems, a substantial fraction of the power of
the triboluminescence emission produced at a remote BBH merger is transmitted into the secondary $\gamma$-rays of the energy
range from 1~GeV to 1~TeV and hence to observe this phenomenon one should rely on the~{\it Fermi}-LAT~\cite{Fermi-LAT}
and very high energy atmospheric facilities, MAGIC~\cite{MAGIC}, HESS~\cite{HESS}, VERITAS~\cite{VERITAS},
HAWC~\cite{HAWC} and LHASSO~\cite{LHAASO}.

Since the electron is massive the arrival timing of the secondary $\gamma$-rays should be delayed by amount of
\begin{equation} \label{teLAG1}
\Delta t(E_e)\simeq\frac{D_{\rm IC}}{c}\frac{m_e^2}{2E_e^2}~,
\end{equation}
where the velocity of light $c$ is introduced explicitly. Thus, for $\simeq$~GeV secondary $\gamma$-rays, which are produced by
$\simeq$~TeV electrons (\ref{EIC1}) covering the Compton scattering distance (\ref{DIC1}) of about $D_{\rm IC(TeV)}\simeq 600$~kpc,
the delay (\ref{teLAG1}) can be as large as $\Delta t({\rm 1\ TeV})\approx 4$~s, which exceeds essentially the duration of the
original signal $t_{\rm obs}=(1+z)t_{\rm mrg}$, in the observer's frame for a BBH at $z\lesssim 1$. In particular, it seems
that the very short pulse of VHE $\gamma$-rays of Hawking radiation from sources at redshift
over $z_{\rm C}\approx 0.2$ (at luminosity distance over 1~Gpc), will be converted into a burst of
$\simeq$~GeV to $\simeq 10$~TeV $\gamma$-rays of $\lesssim 4$~s duration.

The Hawking luminosity,
\begin{equation} \label{bhL1}
L_{\mu\rm BH}\simeq 10^{29}\left(\frac{10^{8}{\rm g}}{M_{\mu{\rm BH}}}\right)^2\ {\rm erg/s}
\end{equation}
implies that a $\mu$BH releases the energy in amount of
\begin{equation}
\label{bhE1}
E_{\mu\rm BH}\simeq 10^{28}\left(\frac{M_{\mu{\rm BH}}}{10^{8}{\rm g}}\right)\ {\rm erg}
\end{equation}
within its evaporation time (\ref{bht1}).

For further estimates let us assume that the total energy release, for the considered effect, at least, could be comparable with the
isotropic equivalent energy of SGRBs, which is of the order of $10^{49}-10^{51}$~erg. Thus, to emit $E_{\rm mrg}\approx 10^{49}$~erg
of the isotropic equivalent energy, in the process of the triboluminescence one would need to evaporate about
\begin{equation} \label{Nmrg1}
N_{\mu\rm BHs}^{\rm SGRB}\simeq 10^{21}
\end{equation}
of $\mu$BHs of mass $M_{\mu{\rm BH}}\approx 5\times 10^7$~g, so that the cumulative mass of the evaporated $\mu$BHs capable to
provide the radiation power similar to a SGRB amounts~$M_{\mu{\rm BHs}}^{\rm tot}\simeq 10^{28}\ {\rm g}\approx 10^{-5}M_{\odot}$.
The value (\ref{Nmrg1}) may serve as a lower estimate of the amount of $\mu$BHs capable to provide an observable electromagnetic
counterpart of the phenomenon. The gravitation waves energy radiated in BBH mergers discovered by LIGO and VIRGO ranges
from few to several solar masses~\cite{bbhLIGO1, bbhLIGO2} and is estimated with precision of about 10\% (see Table III in~\cite{bbhLIGO2}).
It is reasonable to accept as the maximal
possible energy budget of the electromagnetic messenger the amount $0.1M_{\odot}$ ($E_{\rm mrg}\approx 10^{53}$~erg), which implies a generation of
\begin{equation} \label{Nmrg2}
N_{\mu\rm BHs}^{10\% M_{\odot}}\simeq 10^{24}
\end{equation}
in a BBH merger.

Along with $\gamma$ rays, the $\mu$BHs will emit the same amount of isotropic energy equivalent,
\beq
\label{Etribo1}
E_{\rm tribo}\simeq 10^{49}\div 10^{53}\ {\rm erg}~,
\eeq
in neutrinos of the mean energy, observed by an external observer, about 100~TeV, in the process of Hawking evaporation.
Therefore, one expects that the phenomenon should be manifested in arrival of about 10~ms long high energy neutrino
signal in temporal and directional coincidence with the GWs signal from a BBH merger.
The energy release in such neutrino burst can be compared with that one of
the neutrino flare~\cite{IceCubeFlare} arrived from the direction of the blazar TXS~0506+056
prior to the first multi-messenger event in the neutrino-photon astronomy,
baptized IceCube-170922A~\cite{IceCubeMulti}. The energy fluence of the TXS~0506+056 flare
implies that the average isotropic neutrino luminosity delivered by the source
during 158 days is~\cite{IceCubeFlare}
\beq
\label{lumNu1}
L_{\nu}=1.2^{+0.6}_{-0.4}\times 10^{47}\ {\rm erg\ s^{-1} }~.
\eeq
This luminosity is at least two orders of magnitude lower than that one which could
be provided by the triboluminescence phenomenon,
\beq
\label{lumNutr1}
L_{\rm\nu}^{\rm tribo}\simeq\frac{E_{\rm tribo}}{\Delta t_{\rm mrg}}\approx 10^{51}\div\ 10^{55} {\rm erg\ s^{-1} }~.
\eeq
The TXS~0506+056 flare mostly consists of neutrinos with
energies $\gtrsim 20$~GeV, which is an order of magnitude lower than
the energy of neutrinos emitted by the $\mu$BHs. Therefore, the number of
neutrinos in the burst which might arrive from the BBH merger should be at least
one order of magnitude higher than that one in the TXS~0506+056
flare~\cite{IceCubeFlare}. We notice that the energy of IceCube-170922A~\cite{IceCubeMulti}
is reported to be $290$~TeV, which is similar to the average neutrino energy
expected from the $\mu$BHs created in a BBH merger. Therefore, one may expect
that IceCube will be able to detect a neutrino burst from the
multi-messenger effect of the triboluminescence.


\section{Standard model effective potential}
\label{sec:SMvacuumSt}

To fix notations we start from an explicit renormalizable Lagrangian that leads to a vacuum expectation value for the Higgs field $H$
\begin{equation} \label{Higgs}
L = |D_\nu H|^2 - \lambda \left(|H|^2 - v^2\right)~,
\end{equation}
where the low energy self-coupling constant and vacuum expectation value (VEV) are,
\begin{equation}
\lambda \approx 0.13~, \qquad v \approx 246~{\rm GeV}~.
\end{equation}
The expansion of the complex Higgs doublet,
\begin{equation}
H = \frac {1}{\sqrt 2}
\begin{pmatrix}
0\\
v + h
\end{pmatrix}~,
\end{equation}
generates a canonically normalized physical Higgs scalar $h$ of the mass
\begin{equation}
M_h = \sqrt {2\lambda} v \approx 125~{\rm GeV}~,
\end{equation}
with the potential
\begin{equation} \label{U_0}
U_0  = \frac {M_h^2}{2}h^2 + \lambda v h^3 + \frac {\lambda}{4} h^4~,
\end{equation}
which has a minimum $U_0 = 0$ at $h = 0$. In this way, the Higgs doublet develops a non-zero VEV, so that $SU(2)_L\times U(1)_Y$ is broken. The VEV $v=246$~GeV corresponds to the EW vacuum, which is stable at tree level. However, the Higgs potential obtains radiative corrections, so that coupling constants must be running when the energy regime, or correspondingly, the field value, changes.

In quantum field theory, in Minkowski space, the standard regularization \cite{Col-Wei} appears from specific integrals, e.g. for a scalar particle of mass $m$ \cite{Peskin_Book, Schwartz:2013pla},
\begin{equation} \label{int}
\int \frac {dEd^3p}{(2\pi)^4} \frac {1}{\left(E^2 - p^2 - m^2 + i\varepsilon\right)^2} \to - \frac {i}{16\pi^2} \ln \frac{m^2}{\Lambda^2}~,
\end{equation}
where $\Lambda$ is a regularization scale. This implies that an observable, at energy scale $\sim m$, differs from its value at higher energy scale $\sim \Lambda$ by a logarithmic term. Therefore, when quantized,  the Higgs field potential (\ref{U_0}) becomes modified by corrections like (\ref{int}), so that
\begin{equation} \label{U_SM}
U_{\rm eff}(\Lambda) = U_0  + U_1(\Lambda)~,
\end{equation}
with radiative corrections term expressed in the general form as
\begin{equation} \label{U_1}
U_1(\Lambda) = \sum_i \frac {n_i}{64\pi^2}M_i^4 \left[\ln \left(\frac {M^2_i}{\Lambda^2}\right) - C_i\right]~.
\end{equation}
In (\ref{U_1}), the index $i$ runs over particle species, $n_i$ counts degrees of freedom (with a minus sign for fermions), and the field-dependent mass squared of $i$-th specie is given by
\begin{equation}
M^2_i (h) = \Lambda^2_i + g_i h^2~,
\end{equation}
where $g_i$ are coupling constants, while $C_i$ are some definite constants \cite{Schwartz:2013pla}.

For the regularization scale $\Lambda$, in (\ref{U_1}), the VEV of the Higgs field can be chosen as $v$, since this is the scale where the running SM parameters
still can be associated with experimentally observed values, so that the Higgs potential is to be well approximated by its classical form.
It is known, that the leading terms of the one-loop corrections (corresponding to the $t$ quark, $W$ and $Z$ bosons, Higgs and
Goldstone bosons, respectively) are~\cite{smInstable1, smInstable2, smInstable3, smInstable4, smInstable5},
\begin{equation} \label{CW}
\begin{split}
U_1 (v) =& \frac {1}{64\pi^2} \left\{ - 12 \left( \frac 12 Y_t^2 h^2 \right)^2 \left[\ln \frac {Y_t^2 h^2}{2v^2}- \frac 32\right] + \right. \\
&+ 6 \left(\frac {g_2^2}{4}h^2\right)^2 \left[\ln \frac {g_2^2h^2}{4v^2} - \frac 56\right] + 3\left(\frac {g_1^2 + g_2^2}{4} h^2\right)^2 \left[\ln \frac {(g_1^2 + g_2^2)h^2}{4v^2}- \frac 56\right] + \\
&\left. + \left(3\lambda h^2 - \mu^2\right)^2 \left[\ln \frac {3\lambda h^2 - \mu^2}{v^2} - \frac 32\right] + 3 \left(\lambda h^2 - \mu^2\right)^2 \left[\ln \frac {\lambda h^2-\mu^2}{v^2}- \frac 32\right] \right \}~,
\end{split}
\end{equation}
where $\mu^2 = M^2_h/2$ is the Higgs potential mass constant and the $t$-quark Yukawa coupling and gauge bosons coupling constants have the values:
\begin{equation}
Y_t = \frac {\sqrt 2 M_t}{v} \approx 0.99~, \qquad \frac {g_2}{2} = \frac {M_W}{v} \approx 0.33~, \qquad \frac {\sqrt{g_1^2 + g_2^2}}{2} = \frac {M_Z}{v} \approx 0.37~.
\end{equation}
The factor of 12 in the $t$-quark contribution in (\ref{CW}) (first line), corresponds to the 3 colors times 4 components of a Dirac spinor and the minus sign reflects the Fermi statistics. The factor of 3 in the vector boson terms (second line) comes from tracing the numerator of the gauge-boson propagator in the Landau gauge and additional factor 2 in the first term of the second line appears due to the existence of two $W$-boson species.

Below, we review the behavior of $U_{\rm eff}(h)$ for different values of $h$.

\underline{EW regime.} For the EW regime, namely at $h \simeq v$, logarithm terms in (\ref{CW}) are smaller than corresponding constants
$C_i$ in square brackets. The largest contribution comes from the $t$-quark, which appears to be positive making the entire correction
$U_1$ to the Higgs tree level potential $U_0$ also positive. Therefore, at EW scales, radiative corrections do not destabilize the Higgs vacuum which is located at $U_0 (h = 0) = 0$.

\underline{Intermediate regime.} The one loop approximation (\ref{CW}) is still valid at intermediate scales,
$h \simeq 10 v \simeq 1$~TeV. At this regime logarithm terms in (\ref{CW}) become positive and larger than constants
$C_i$, so that the effective potential (\ref{U_SM}) is well approximated by
\begin{equation} \label{U1}
U_{\rm eff}(h)  \approx \left[\frac {\lambda}{4} + \frac {3}{64\pi^2} \left( 3\lambda^2 - Y_t^4 \right) \ln \frac {h^2}{v^2}\right] h^4~.
\end{equation}
Due to the minus sign in front of the $t$-quark term the one loop correction becomes negative, i.e. the SM effective
potential has a local minimum. However, due to the large value of the denominator, $64\pi^2$, the second term
in (\ref{U1}) is smaller than $\lambda/4$, so that the effective potential remains positive, i.e. higher than
the vacuum value $U_0 = 0$. Therefore, in Minkowski space, at the scales $\sim 1$~TeV, the global minimum of the Higgs system is still located at $h=0$.

\underline{Large Higgs field regime.} For larger values of the Higgs field $h >>$~1 TeV
the effective potential (\ref{U_SM}) can be approximated~\cite{cosmology2, higgsStrumia1, higgsStrumia2} as
\begin{equation} \label{U_h}
U_{\rm eff}(h) \approx \frac {\lambda(h)}{4} h^4~.
\end{equation}
where $\lambda(h)$ is the running coupling.
The behavior of the running coupling $\lambda(h)$ in the renormalization group approach is shown in Fig.~\ref{fig:Lambda}.
\begin{figure}[h]
\centering
\includegraphics[width=0.5\textwidth]{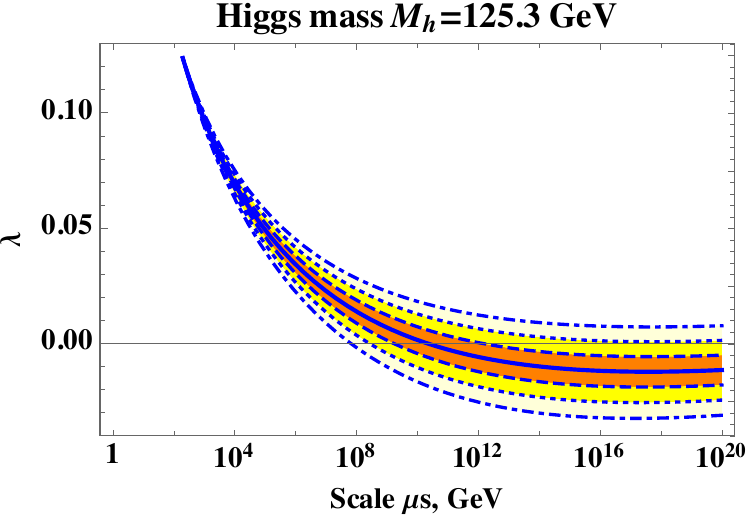}
\caption{Higgs self-coupling $\lambda$, obtained in the framework of $\overline{MS}$
renormalization scheme for central values $M_h = 125.3$~GeV and $M_t = 172.76$~GeV.
Deviations for $M_t\pm 1$~GeV, $M_t\pm 2$~GeV, $M_t\pm 3$~GeV are shown.}
\label{fig:Lambda}
\end{figure}
The renormalisation group evolution of the running coupling demonstrates that
$\lambda(h)$ becomes negative above $10^8$~GeV  and is very sensitive to the top quark mass~\cite{fedor, gkr}.
Therefore, in general, SM vacuum is a local minimum and it becomes unstable at energies much higher than the EW scale.


\section{Gravitational Corrections to Vacuum Decay}
\label{sec:GravCorr}

Although the vacuum decay rate in Minkowski space is extremely slow, that is not necessarily the case
in strong gravitational fields. The full theory of quantum gravity is unknown and only exists a simple
semi-analytic approximation that captures the leading gravitational corrections~\cite{Isidori:2007vm, Salvio:2016mvj}.
It was found~\cite{Hiscock:1987hn, Berezin:1990qs, hBH1, hBH2, hBH3, hBH4, hBH5, hBH6} that due to
external gravitational field the vacuum decay can be significantly increased. This raises the
question of SM vacuum stability in external classical gravitational potential. For example, close to the
horizon of a BH the strong local spacetime curvature can enhance vacuum decay rate to a level incompatible
with the age of the Universe \cite{hBH2, hBH3, hBH4}.

Existing calculations on gravity mediated Higgs vacuum decay mainly consider small BH's located inside new
phase bubbles and are taking into account only the effects of
curvature on the running of the SM constants. These corrections appear after inserting the non-minimal
coupling term, $\sim |H|^2 R$, into the action that connects the Higgs field $H$ to the scalar curvature of gravity $R$.

In this paper we want to estimate the impact of gravitational potential on the rate of the nucleation of
the new phase bubble close to the horizon of a BH of astrophysical origin. Let us ask the important question: Is the SM vacuum
stable in large gravitational potential? The models of quantum theory in curved spacetime is a topic of intense
research~\cite{Bir-Dav, Wald}, which revealed many interesting phenomena such as the Hawking radiation and the Unruh effect, etc.
Here we will consider modifications of the dispersion relations for quantum fields in the renormalization
integrals (\ref{int}) in external large static gravitational potential. Similar ideas were considered in
models of quantum theory in cosmological spaces, where dispersion relation in the integrals (\ref{int}) is
modified by inserting of the scale factor $a(t)$ describing cosmic expansion (see the recent review~\cite{Markkanen:2018pdo}).

Note that the gravitational deformation of dispersion relations assumes an effective violation of the Lorentz invariance.
Several models predict a departure from exact Lorentz invariance \cite{LIV}, when the free particle dispersion relation
exhibits extra momentum dependent terms, apart from the usual quadratic one occurring in the Lorentz invariant
dispersion relation. Most of the studies were performed in the QED, gravity, and for some of the SM particles,
where strong constraints on the Lorentz invariance violating terms were obtained \cite{Mattingly:2005re}. However,
not much studies were done for the Higgs sector~\cite{Higgs}, especially on curved backgrounds. Higgs is much different
from other fundamental fields and one cannot exclude a possibility that the large effective Lorentz violation,
close to the horizon of a BH, could be important only in the Higgs sector.

Usually, it is assumed that close to the horizon of an astrophysical BH, where
gravitational invariants are negligible, a curvature has a small impact on quantum particles.
In many cases the most useful invariant is the Kretschmann scalar~\cite{gravInv1},
which is inversely proportional to the sixth power of the radius of a BH.
However, quantum vacuum is not empty Minkowski space and a
large gravitational potential can significantly change renormalization parameters. It is known that the density
of particles states at large energy grows exponentially \cite{hagedorn}. So, if a gravitational energy is pumped
into the system, new higher mass states are produced (rather than that the energy of already existing states is
increased, which implies an increase in the temperature), which can amplify loop corrections in quantum vacuum.

In general, if quantum fluctuations in the Higgs sector create a bubble of a new phase that is large enough, it is
energetically favorable for this bubble to expand due to the gain in volume energy over the energy stored in the bubble wall.
On the other hand, if the bubble is small its surface tension compresses the bubble and it disappears.
The addition of an external gravitational potential, i.e. an additional energy, can change the situation dramatically.
The distortion of space changes the balance between volume and surface energies, the ``cost'' of bubbles formation is lower
and the bubbles with smaller radii can be created.


\section{Quantum vacuum at a horizon}
\label{sec:QvacuumH}

Let us estimate the impact of the large gravitational potential on the properties of a quantum vacuum close to the horizon of a BH of mass $M$. For simplicity, we write the Schwarzschild metric in isotropic coordinates,
\begin{equation} \label{metric}
ds^2 = V^2(R) dt^2 + A^2(R) \delta_{ij}dx^idx^j ~, \qquad (i, j = 1,2,3)
\end{equation}
where the metric coefficients are
\begin{equation}
V(r) = \frac {1 - M/2R}{1 + M/2R}~, \qquad  A(r) = \left(1 + \frac {M}{2R}\right)^2~.
\end{equation}
Here the radial function $R^2 = \delta_{ij}x^i x^j$ has the following expression in terms of the Schwarzschild radial coordinate $r$:
\begin{equation} \label{2R}
2R = r-M + \sqrt {r^2 - 2Mr}~,
\end{equation}
which only holds if $R \to \infty$ when $r \to \infty$, and outside the event horizon, $r \geq 2M$. It follows from (\ref{2R}) that close to the horizon $r \to 2M$, we have $2R \to M$ and $V(r) \to 0$ and, unlike the Schwarzschild case, the isotropic metric (\ref{metric}) leads to a real singularity at the Schwarzschild horizon, since the determinant
\begin{equation} \label{det}
\sqrt{-g} = V A^3 =  \left(1 - \frac {M}{2R}\right) \left(1 + \frac {M}{2R}\right)^5 ~,
\end{equation}
becomes zero at $R = M/2$, or at $r = 2M$.

Since $V^2 \leq 1$ and $A^2 \geq 1$, for a distant observer the gravitational potential $M/2R$ effectively reduces the speed of light and
increases spatial distances close to the BH horizon. This will reduce the volume of integration,
\begin{equation}
d^4x \to |VA^3|d^4x = \left(1 - \frac {M}{2R}\right) \left(1 + \frac {M}{2R}\right)^5d^4x~,
\end{equation}
and thus will modify the Euclidean action with an external gravitational potential for the bounce solution,
\begin{equation}
\label{Smod}
S_4 \to \left(1 - \frac {M}{2R}\right) \left(1 + \frac {M}{2R}\right)^5~S_4~.
\end{equation}
So, the bounce has a smaller action and the vacuum decay process can be significantly enhanced for the large gravitational potentials ($M/2R \to 1$) at some proper distance from the Schwarzschild horizon.

As an example let us consider a scalar particle of mass $m$ on the Schwarzschild background. The radial geodesic equation has the form:
\begin{equation} \label{geodesic}
\frac {E^2}{1 - 2M/r} - \left(1 - \frac {2M}{r}\right) p_r^2 - m^2 = 0~.
\end{equation}
Note that close to the horizon $r \to 2M$, or for the relativistic case $p_r \gg m$, the mass term in (\ref{geodesic}) can be neglected.
It is known that the energy of a relativistic particle and hence its frequency in static gravitational potential does not depend on the
distance from the gravitating body~\cite{Okun}. As the energy $E$ does not depend on $r$, we immediately obtain from (\ref{geodesic})
that its momentum $p_r$ does depend and in the $m \to 0$ limit
\begin{equation} \label{p_r}
p_r \approx \frac {E}{1 - 2M/r}~.
\end{equation}
So, the closer is the particle to the horizon, the larger is its
momentum~\footnote{The example is considered in order to stress the point that the energy of a particle does not depend on gravitational potential $M/2R$,
so that, in the dispersion relation (\ref{geodesic}), the momentum and the mass terms are affected by $V$. Also, at the
horizon $V \to 0$, which implies that $V \cdot m \to 0$ and hence the mass term does not destroy the renormalization scheme.}.
In the isotropic coordinates (\ref{metric}) with the constant parameters $V$ and $A$ the dispersion relation (\ref{geodesic}) takes the form
\begin{equation} \label{dispersion}
E^2 - P^2 - V^2 m^2 = 0~, \qquad \left(P^2 =  \frac{V^2}{A^2}p^2\right)
\end{equation}
and the integral (\ref{int}) is modified as
\begin{equation}
\label{hMod1}
\frac {A^3}{V}\int \frac {dEd^3P}{(2\pi)^4} \frac {1}{\left(E^2 - P^2 - V^2 m^2 + i\varepsilon\right)^2} \to - \frac {iA^3}{16V\pi^2} \ln \frac {m^2}{\Lambda^2}~.
\end{equation}
One may comment that in static gravity, to preserve dispersion relations with constant energy, the momentum and mass terms should be modified. Then in Pauli-Villars regularization, for example, we have a real and the ghost particle, with the mass $m$ and some fixed large mass $\Lambda \gg m$, both affected by $V$,
\begin{equation}
\int \frac {dEd^3P}{(2\pi)^4} \left[\frac {1}{\left(E^2 - P^2 - V^2 m^2 + i\varepsilon\right)^2} - \frac {1}{\left(E^2 - P^2 - V^2\Lambda^2 + i\varepsilon\right)^2}\right] \to - \frac {i}{16\pi^2} \ln \frac {m^2}{\Lambda^2}~,\nonumber
\end{equation}
so that the factor $V$ does not show up in the logarithm.

Similar results can be obtained for vector particles. The definition of the fermion propagators summing is performed by tetrads instead of
metric. However, the factor containing the gravitational potential for the fermionic loop contribution, in (\ref{U_SM}),
is the same as for scalars and vectors, since the trace of an odd number of Dirac $\gamma$-matrices is zero and we must
consider an even number of fermion propagators. Then the calculation becomes exactly the same as for the scalar and
vector cases except the important difference of an overall minus sign (due to the Fermi statistics) for the fermionic loop integral.

Thus, we conclude that close to the BH horizon the entire radiative corrections in (\ref{U_SM}) will be modified by the universal factor $A^3/V$,
\begin{equation}
\label{hMod2}
U_1 \to \frac {A^3}{V}U_1~.
\end{equation}

In summary, within the considered scenario,
integrals of the type presented in equation (\ref{int})
 are subject to modifications, as demonstrated by equation (\ref{hMod1}). Therefore, in the
vicinity of a BH horizon, the standard radiative corrections described by equation (\ref{U_SM})
are further influenced by equation (\ref{hMod2}).
This implies that near the horizon, the one-loop
corrections (\ref{CW}) are altered in a
manner that shifts the position of
the maximum of the Higgs effective potential closer to the electroweak scale. As a result, the vacuum
destabilization and the possibility of nucleating new phase bubbles become feasible at significantly lower scales.

Contrary to the seemingly natural assumption that an astrophysical BH cannot possess a large enough curvature, even at the horizon, to be able to
influence the regularization discussed above, one can elaborate as follows.
For the case of the Schwarzschild solution, the origin point is considered as a true physical singularity,
which appears in quantities that are independent of the choice of coordinates, like the Kretschmann scalar. In contrary, the singularity at the horizon is called a
coordinate singularity, which can be avoided by changing to ``good'' coordinates.
However, the necessary ingredient of all these singular coordinates is the so called Regge-Wheeler's
tortoise coordinate, which does not belong to the $C^2$-class of admissible coordinate transformations.
Then, the singular transformations (like introduced by Kruskal-Szekeres, Eddington-Finkelstein, Lemaitre, or Gullstrand-Painleve)
give delta-functions in the second derivatives (see, for example~\cite{gravMTW}, for details). This means that transformed metric tensors at the horizon are not differentiable,
i.e. they are of unacceptable class $C^0$. The Einstein equation for these metrics is altered with fictitious delta-sources
at the horizon~\footnote{Recall that the $C^2$-differentiability assumption for metric tensors plays a key role in the singularity theorems of
the general relativity. Thus the second order partial derivatives of the metric tensor should exist and be continuous.
The initial value analysis also leads to the restriction that admissible coordinate transformations are to be of class $C^2$,
since they should not change the Riemann tensor, which would lead to the fictitious extra source in the Einstein equations.}.
For a sufficiently large BH, one can let Kretschmann invariant to be arbitrarily small at the horizon.
This would be usually interpreted as we had not a large curvature, so that one can use Minkowski space to describe particles.
However, the conclusion on a finiteness of the Kretschmann invariant, at the
horizon, is usually based on an assumption of a mutual cancellations of delta-function source like divergences.
The same is true for other invariants of the
gravitational field. In general, metric components are independent functions and the cancellation of their
zeros at the horizon is accidental, since it follows from the exact validity of the vacuum Einstein equations
implying a perfect sphericity. However, a perfect spherical symmetry and true vacuums are rarely observed, if ever.
Therefore, a smallness of the Kretschmann scalar does not mean that curvature is small at the BH horizon.
Indeed, the three from six non-zero independent components of the mixed Riemann tensor
for Schwarzschild metric blow up at the horizon. So, we consider a model where at the BH horizon
space-time is not Minkowskian (see, for example~\cite{meG1,meG2,meG3,meG4}).
In summary, the extension of geodesics across the Schwarzschild horizon by singular diffeomorphisms
presents difficulties even at the classical GR level. The Regge-Wheeler radial variable
expression should include a Heaviside function at the Schwarzschild horizon,
which corresponds to delta-like sources and leads to infinite Riemann and Ricci tensors~\cite{meG4}.
Thus, plane wave solutions that cross the horizon are not viable, and we must establish
appropriate boundary conditions and redefine the concept
of the quantum vacuum in proximity to the Schwarzschild horizon~\cite{meG4}.


\section{The toy model}
\label{sec:ToyM}

In order to estimate a vacuum decay rate and new phase bubble parameters let us consider a toy model presenting the effective potential (\ref{U1}) in the form
\begin{equation}\label{pot}
\begin{split}
U(h) \approx \frac{\lambda}{4}h^4 \left[1 - \frac{2}{k} \ln \frac{h}{v}\right]~.
\end{split}
\end{equation}
In (\ref{pot}) $k$ parameter is given by
\begin{equation} \label{k}
k = \frac{16 \pi^2 \lambda ~V(R)}{3 (Y_t^4 - 3\lambda^2)~ A(R)^3} \approx 9.3  \frac{V(R)}{A(R)^3} \approx 0.073d_{\rm H} \ll 1~,
\end{equation}
where couplings were defined at $M_t$ scale as $Y_t \approx$~0.934 and $\lambda \approx$~0.127 and changes in these couplings were neglected up to a few TeV scale.
It follows from (\ref{pot}) that in Minkowski space-time, where $V = A = 1$, the SM vacuum at $h = 0$ is meta-stable but the probability of its decay is extremely low.
However, in a strong gravitational field close to a BH horizon the parameter (\ref{k}) decreases and at
some distance to the horizon the potential (\ref{pot}) becomes negative already at $h \approx v$
leading to a significant vacuum instability. The potential  (\ref{pot}) is shown in Fig.~\ref{fig:U_no_Grav} at $k=0.1$.

\begin{figure}
\centering
\includegraphics[angle=0,scale=1.0]{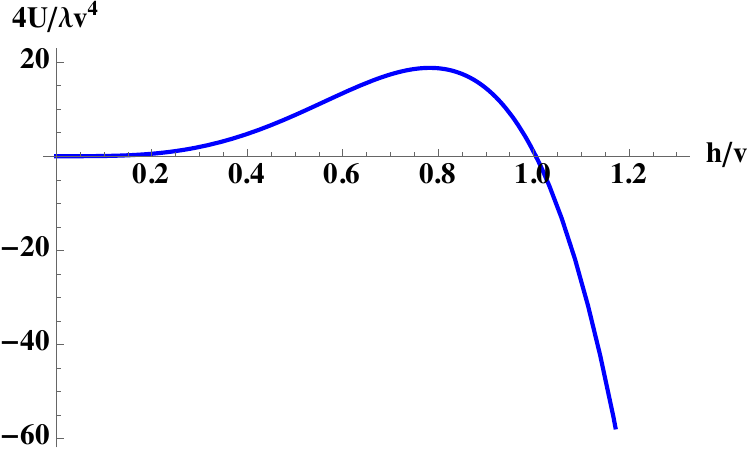}
\vspace{0.5cm}
\caption{\it The modified Higgs potential (\ref{pot}) at $k=0.1$.}
\label{fig:U_no_Grav}
\end{figure}

The probability of vacuum decay can be calculated by solving numerically the equation of bounce motion.
The presence of a BH breaks the translational symmetry and as the result a bounce shape corresponding to
the minimal action can be not spherically symmetric. However, a solution with  $\mathcal{O}(4)$ symmetry can be used as the lower bound on the vacuum decay probability.

The $\mathcal{O}(4)$ bounce solution \cite{Coleman:1977} is a configuration satisfying the Euclidean field equation
\begin{equation}\label{eom}
\frac{\partial^2 h}{\partial \rho^2}+\frac{3}{\rho} \frac{\partial h}{\partial\rho} = \frac{\partial U(h)}{\partial h}~,
\end{equation}
with the boundary condition $U \rightarrow 0$ (false vacuum) as $\rho \rightarrow \infty$~, where $\rho$ is the Euclidean 4-radius
\begin{equation}
\rho = \sqrt{t^2 + x_{i}x^i}~. \qquad (i = 1, 2, 3)
\end{equation}

\begin{figure}
\centering
\includegraphics[angle=0,scale=1.0]{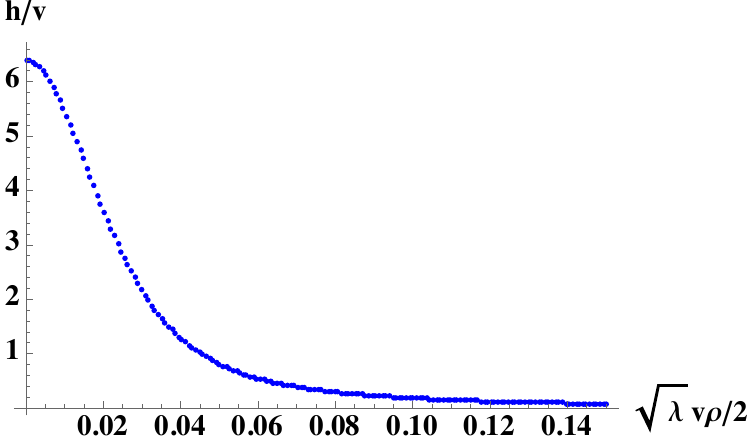}
\vspace{0.5cm}
\caption{\it The bounce solution of Eq.~(\ref{eom}) at $k=0.1$.}
\label{fig:Bounce}
\end{figure}

The equation (\ref{eom}) is not exactly solvable. We solve it numerically by using $\textit SimpleBounce$, a $C++$ package \cite{Sato:2020} for finding the bounce solution for the false vacuum decay. An example of a numerical solution with $k = 0.1$ obtained in an assumption that a distance to the Schwarzschild sphere is constant during nucleation and collapse of the new phase bubble presented in Fig.~\ref{fig:Bounce}. For different values of $k$ the bounce solution can be parameterized as
\begin{equation}\label{sol}
h(\rho) \approx \frac{6.36v} {\sqrt{1 + \bigg({\frac{\lambda}{2k}}\bigg)^2 \bigg(b v \rho \bigg)^4}}
\end{equation}
where $b \approx 5.9$ and the critical radius $\rho_{c}$ of a true vacuum bubble is given by
\begin{equation}\label{rhoc1}
\rho_{c} \approx \frac{0.045}{\sqrt{\lambda} v} \sqrt{\frac{k}{0.01}} \approx 10^{-17} \sqrt{\frac{k}{0.01}} {\rm cm}~.
\end{equation}
At high nucleation rate, one distinguishes expanding and contracting bubbles. In order for the bubbles to expand rather than contract, it is necessary that the gain in volume energy from the bubble interior being in the lower free energy phase overcomes the unfavorable surface tension of the bubble. This will happen if the bubble radius is larger
than the critical radius $\rho_c$, for which the two effects balance out~\cite{Coleman:1977, bubbleOkun1, kolbSub-critical}.

For the $\mathcal{O}(4)$ bounce the Euclidean action is given by the equation:
\begin{equation}\label{action}
S_{4} = 2 \pi^2 \int d\rho \ \rho^3 \left[ \frac{1}{2}\bigg(\frac{d h}{d\rho}\bigg)^2 + U(h)\right]~.
\end{equation}
This equation, in combination with (\ref{Smod}), gives the action for the critical bubble in a gravitational field
\begin{equation}
\label{rhoc}
S_{4} \approx 0.0645 \frac{k}{0.01} \frac{128}{\lambda} d_{\rm H} \approx 480d_{\rm H} ^2~.
\end{equation}

\section{Discussion}
 \label{sec:disc}
The toy realization, described above, implies that the modification of the Higgs vacuum decay in the vicinity of a BH corresponds
 to $A_S\approx 480$ and $a=2$ introduced in heuristic formula (\ref{effAct1}).
Besides this, as it follows from Fig.~\ref{fig:U_no_Grav}, the Higgs potential is modified in a way
that the position of its maximum is moved close to the values of the EW scale. In its turn, the bounce solution, presented in Fig.~\ref{fig:Bounce}, indicates that
\beq
\label{hb1}
h_{\rm b}\approx 6v\approx 1.5\ {\rm TeV}~,
\eeq
in (\ref{tl2}). Therefore, according to (\ref{tl3}), bubbles which are capable to convert into a $\mu$BHs of
mass $M_{\mu{\rm BH}}\approx 5\times 10^7$~g by the Higgs vacuum decay at $h_{\rm b}\approx 1.5$~TeV should have
radius $R_{\rm bub}\simeq 9\times 10^{-4}$~cm. This implies that the collision (percolation) time
$\Delta t_{\rm col}\simeq 30$~fs is negligible being compared with the merging time $t_{\rm mrg}\simeq 10$~ms, which ensures that in the case
of maximal effective triboluminescence discussed in Section~\ref{sec:emSignal}, the $\mu$BHs of
required masses will be formed almost immediately to have enough time to evaporate out completely
and produce the detectable gamma ray and neutrino signals with characteristics elucidated in Section~\ref{sec:emSignal}.

Notice, the temporal properties of disturbed space-time dynamics at the merging
are also defined by $t_{\rm mrg}$. Since it takes a negligible fraction of the merging
time to grow the bubbles to the relevant size, any dynamical evolution of the space-time within UVS,
during the bubbles conversion into $\mu$BHs, should not affect the formation of $\mu$BHs.
Also, it is assumed that the Hawking evaporation of $\mu$BHs, immersed into rapidly evolving space-time
within the gap of the merger, stays intact.

One can express the total numbers of $\mu$BHs (\ref{nmbh1}), (\ref{nmbh2})  created by $s=2$ and  $s=3$ mechanisms,
in the first scenario allocated in section \ref{sec:sandw},
as follows
\begin{equation}
\label{nmbh12}
N_{\mu{\rm BHs}}(d_{\rm H},2)\approx 5.5\times 10^{174}d_H^6\kappa_f^{1/2}{\cal P}_{\mu{\rm BH}}(2)
\langle \Gamma (d_H) \rangle^2 \left(\frac{1}{\rm 1\ GeV^8}\right)~,
\end{equation}
\beq
\label{nmbh13}
N_{\mu{\rm BHs}}(d_{\rm H},3)\approx 4.3\times 10^{247}d_H^8{\cal P}_{\mu{\rm BH}}(3)
\langle \Gamma (d_H) \rangle^3 \left(\frac{1}{\rm 1\ GeV^{12}}\right)~.
\eeq
These estimates are obtained for a BBH merger with components of mass $10M_{\odot}$.
Using the numerically computed action (\ref{rhoc})
along with (\ref{PhTrR1}) and (\ref{effAct1}) one can express the average decay rate as follows
\begin{equation} \label{RS1}
\langle \Gamma (d_{\rm H})\rangle \approx{\cal M}^4\exp{[-480d_{\rm H}^2]}~.
\end{equation}
Some examples of analytical and numerical calculations of the
pre-factor ${\cal M}$ can be found in  ~\cite{tmf, amendola, dunne, guada}.
In the following ${\cal M}=v$ will be used as a rough value in (\ref{nmbh12}) and (\ref{nmbh13}).

Comparing the estimates (\ref{nmbh12}) and (\ref{nmbh13}) with the number of $\mu$BHs  needed to provide
the energy budget of detectable electromagnetic counterpart, as discussed in Section~\ref{sec:emSignal},
we arrive to the following estimate
\begin{equation}
\label{dh2}
d_{\rm H}\lesssim 0.64~,
\end{equation}
made for $\kappa_f\approx 10^{-3}$, taken from assessment of~\cite{pbhBubble3}.
The estimate (\ref{dh2}) is valid for both $s=2$ and $s=3$ mechanisms.

Due to the logarithmic
dependence defining the value $d_H$ of (\ref{dh2}), $d_H$ is quite weakly sensitive
to $\kappa_f$ and ${\cal P}_{\mu{\rm BH}}(s)$ and stays almost the same for both
quantities (\ref{Nmrg1}) and (\ref{Nmrg2})~\footnote{Our current ignorance
of details of the mechanism of ${\mu{\rm BHs}}$ formation in bubble collisions
which might be represented by a poor knowledge of the value of ${\cal P}_{\mu{\rm BH}}(s)$, say up to
about 10 orders of magnitude uncertainty ($10^{-10}\lesssim {\cal P}_{\mu{\rm BH}}(s))\lesssim 1$),
results only in about 10\% variation in estimate (\ref{dh2}). A detailed consideration of
the formation of BHs in bubble collisions is in preparation~\cite{pbhBubble6}.}.

Finally, evaluating the condition (\ref{condSc2}) which
corresponds to the second scenario, discussed in section \ref{sec:sandw},
we obtain
\begin{equation}
\label{dh3}
d_{\rm H}\lesssim 0.23~.
\end{equation}

These results, obtained within the framework of the
aforementioned toy model above, substantiate our
conjecture that the nucleation of new phase
bubbles may be enhanced in a region “sandwiched”
between the horizons of merging BHs.

Notice that, in second scenario, in order to produce the $\mu$BHs in amount of $N_{\mu\rm BHs}^{\rm SGRB}\simeq 10^{21}$
and $N_{\mu\rm BHs}^{10\% M_{\odot}}\simeq 10^{24}$, the colliding bubbles need
to occupy the volumes about $3 \times 10^{-3}\ {\rm km}^3$ and $3\ {\rm km}^3$,
respectively. Certainly, these volumes can fit inside $V_{\rm UVS}(d_{\rm H})\approx 1.1\times 10^4\ {\rm km}^3$
calculated from (\ref{vol2}) for $d_{\rm H}=0.23$ in case of $\simeq 10M_{\odot}$
components of BBH merger. In other words, the amount of the Higgs meta-stable
vacuum needed to produce a planetary cumulative mass of
$\mu$BHs, which are capable to provide SGRB energy budget for the electromagnetic
messenger, is about $3 \times 10^{-3}\ {\rm km}^3$. A $3\ {\rm km}^3$ of the Higgs
meta-stable vacuum is capable to emit about 10\% of $M_{\odot}$ in Hawking radiation.

It is expected that a combination of both scenarios,
discussed in section \ref{sec:sandw}, will take place. This implies that
photons and neutrinos are suppose to be emitted within the distance range $0.6\lesssim d_{\rm H}\lesssim 0.2$,
defined by (\ref{dh2}) and (\ref{dh3}), so that the gravitational energy reduction factor (\ref{zgrav3}),
$z_{\rm g}+1\approx 2$. This red shift was used for the model of the electromagnetic and neutrino signals, developed
in section~\ref{sec:emSignal}, for $\simeq 100$~TeV outgoing photons and neutrinos.

It is extremely unlikely that any expanding bubble wall, escaping  from the UVS, would
survive in a way  to be able to trigger the decay of the SM metastable vacuum everywhere.
Indeed, to reach the outer space by its domain wall
an expanding bubble should ultimately
fill in the entire volume of UVS which will contain already a great multitude
of small, while already overcritical bubbles, due the conditions of high nucleation rate created inside
UVS within $t_{\rm mrg}\simeq 10$~ms. Colliding with the multitude of bubbles the expanding wall will be numerously
perforated so that the continuous topology of the wall providing the vacuum decay conditions will be
destroyed leaving only some rapidly decaying peaces of the wall escaping out of the UVS. Moreover,
occupying the entire volume a potentially escaping bubble should ultimately touch the horizons of
the components of a BBH merger, which in their turn still covered by the ongoing process of
bubble nucleation. The collisions with those at horizons bubbles the constituents of the bowling
substance will also lead to the ultimate perforation of the escaping wall and destruction of its
continuous topology.
The bubble with expanding walls, if any, will fill in the entire volume of UVS and collides with the horizons provided that
$V_{\rm escB}=\frac{4}{3}\pi (ct_{\rm mrg})^3$ exceeds $V_{\rm UVS}$ given by (\ref{vol2}) . This
condition can be recast into constraint on the mass of BBH component, which
reads $M_{\rm BH}\lesssim\frac{800M{\odot}}{d_{\rm H}}$. Therefore, it is extremely
unlikely that BBH mergers with parameters considered in this study could trigger
the decay of the global metastable state of the SM vacuum.

Furthermore we note that, it would be interesting exploring the application of the considered model of gravitational
modifications of the Higgs potential to other classes of models with strong gravitational fields,
particularly in the context of particle models in the early universe. These investigations have the
potential to reveal alternative mechanisms for the cancellation of vacuum energy during inflationary
epochs like presented in~\cite{Luongo:2018lgy, DAgostino:2022fcx, Belfiglio:2023eqi}, offering valuable insights
into the interplay between gravity and fundamental particle physics. By extending our understanding
in this direction, we can gain a deeper comprehension of the early universe and its intriguing dynamics.


\section{Conclusions}
\label{concl}

In this paper we explored the phenomenon that may occur in a vicinity of a BH horizon due to the gravitational corrections to the Higgs potential.
The gravitational corrections can provide conditions for the Higgs vacuum decay, so that a BH being immersed into the EW vacuum will be encompassed
by a thin shell consisting of a ``bowling substance'' represented by nucleating bubbles of new vacuum phase surrounded by the SM vacuum.
Since the nucleating bubbles fall under the horizon an external observer of a single BH should always stay in the EW meta-stable state.
However, in the gap between components of a BBH merger, one might expect a formation of a volume with effective zero gravity, so that
for a short time period the unstable vacuum is kind of ``sandwiched'' between the horizons of the components. Within the time of existence
of the ``sandwich'' the bubbles, nucleated in its volume, will collide and convert their energy into $\mu$BHs, which
in their turn will be evaporated by emission of the Hawking radiation within the final stage of the coalescence of the BBH.
The energy release in this burst-like evaporation, at a BBH of LIGO-Virgo scale, may range from the isotropic energy equivalent
of SGRBs up to 10\% fraction of $M_{\odot}$. We call this phenomenon the Higgs induced triboluminescence in an analogy to that one existing in solid state physics
when materials could emit light being mechanically stimulated, such as rubbing, grinding, impact, stretching, and compression.

Due to the emission of very high energy radiation $\gtrsim 100$~TeV of all SM particles and, maybe, beyond SM species the
phenomenon of Higgs induced triboluminescence in BBH mergers should have quite distinguishable multi-messenger signatures.
Indeed, a gravitational wave signal from a BBH merger arriving at currently running LIGO, VIRGO and KAGRA, or to be constructed in
the future, more sensitive, gravitational wave facilities, should be accompanied with a $\gtrsim 100$~TeV neutrino signal
in IceCube detected from the same direction and with an electromagnetic counterpart to be registered by space based
gamma ray monitors~\cite{SWIFT, INTEGRAL, Fermi-GBM}, the high energy gamma ray telescope~\cite{Fermi-LAT} and/or very
high energy atmospheric Cherenkov facilities~\cite{MAGIC,HESS, VERITAS, HAWC, LHAASO}.
Propagating through the universe, gamma rays of such a high energy will develop electromagnetic cascades
due to $e^+e^-$ pair production on cosmic microwave background and extra-galactic background light,
which will multiply the TeV-GeV spectral component at the cost of very high energy thermal Hawking emission.
However, whatever the arrived spectrum of the gamma rays is, we have always to expect a burst-like temporal behavior
of the signal. The observation of this phenomenon involves a detection of three types of messengers, namely gravitational
waves, very high energy neutrinos and gamma rays, which makes it a perfect physics case for the multi-messenger campaign.

In the case of observation of the effect of Higgs induced triboluminescence in BBH mergers one benefits for valuable
impacts on understanding of at least two very intriguing topics of contemporary physics, namely,
the decay of meta-stable state of the EW vacuum and the evaporation of BHs via Hawking radiation.

\section{Acknowledgments}

The work of Mariam Chitishvili was supported by the Shota Rustaveli National Science Foundation of Georgia (SRNSFG) through the grant  PHDF-19-6294.
The work of Rostislav Konoplich was partially supported by the Kakos Endowed Chair in Science Fellowship.

\end{document}